\documentstyle[preprint,epsf]{jpsj}
\def\EQ{\begin{eqnarray}}
\def\EN{\end{eqnarray}}
\def\bv{{\bf v}}
\def\br{{\bf r}}
\def\str{\sigma^{p}}

\def\bvk{{\bf v_{\bf k}}}
\def\bvt{{\bf v_{\bf k}^\bot}}
\def\hW{\hat W}

\def\bfo{{\bf f}}
\def\bfk{{\bf f_{\bf k}}}
\def\bfl{{\bf f^{||}_{\bf k}}}
\def\bft{{\bf f^{\bot}_{\bf k}}}
\def\bk{{\bf k}}

\def\runtitle{Viscoelastic Effect on Hydrodynamic Relaxation}
\def\runauthor{Akira {\sc Furukawa}}

\title
{%
Viscoelastic Effect on Hydrodynamic Relaxation in Polymer Solutions
}

\author
{%
Akira \sc{Furukawa}
}

\inst{%
Department of Physics, Kyoto University, Kyoto 606-8502\\
}

\recdate
{
\today
}

\abst{%
The viscoelastic effect on the hydrodynamic relaxation in semidilute 
polymer solutions is investigated.
From the linearized two-fluid model equations, 
we predict that the dynamical asymmetry coupling 
between the velocity fluctuations and the viscoelastic stress 
influences on the hydrodynamic relaxation process, 
resulting in a wave-number-dependent shear viscosity. 
}

\kword{%
viscoelasticity, dynamical asymmetry coupling, viscoelastic length  
}

\begin{document}
\sloppy
\maketitle

The viscoelasticity in entangled polymer solutions has 
been extensively studied from various viewpoints\cite{de-Gennes,Doi-Edwards}.
When we investigate the mesoscopic dynamics of polymer solutions, 
the two-fluid model\cite{Brochard-deGennes,Brochard-deGennes2,Doi-Onuki} 
gives a good starting point.
The two-fluid model was put forward by Brochard and de Gennes
\cite{Brochard-deGennes,Brochard-deGennes2}. 
Later, the current form of the two-fluid model was derived by Doi and Onuki
\cite{Doi-Onuki}. 
Its validity has been recognized through researches of a number of 
intriguing viscoelastic phenomena such as 
shear-induced concentration fluctuations
\cite{Doi-Onuki,Onuki1,Helfand-Fredrickson,Milner},
nonexponential decay in dynamic scattering near equilibrium
\cite{Brochard-deGennes,Doi-Onuki,Adam-Delsanti},
phase separation dynamics\cite{Tanaka2,Onuki-Taniguchi,Taniguchi-Onuki,Onuki-book}, and so on.
The two-fluid model contains the following important concepts for
understanding the above various phenomena. 
(i) In entangled polymer solutions, the network stress can act on the two 
components asymmetrically. 
That is, in polymer solutions, the stress is supported by the polymer chains
much more than by the solvent, and the resultant imbalance of the stress
causes the relative motion of the two components.
This effect is called dynamical asymmetry coupling
\cite{Brochard-deGennes,Brochard-deGennes2,Doi-Onuki}. 
(ii) Considering the viscoelastic effect on the transport phenomena 
in polymer solutions, the cooperative motion of the network structure
resulting from the entanglement of the polymers is also important.
Such a cooperative scale intrinsic to entangled polymer solutions 
is characterized by  the so-called viscoelastic length $\xi_{ve}$, 
which was first introduced by Brochard and de Gennes
\cite{Brochard-deGennes,Brochard-deGennes2}.

Previously, Doi and Onuki\cite{Doi-Onuki} 
showed that the diffusion process is drastically influenced 
by the dynamical asymmetry coupling between the concentration fluctuations 
and the longitudinal mode of the viscoelastic force
(stress-diffusion coupling).
Their analysis predicted an additional wave number dependence
of the kinetic coefficient for concentration due to the viscoelasticity, 
which was confirmed by recent experimental studies
(see ref. 14 for example).
However, in their analysis, 
very viscous systems without the velocity fluctuations were assumed, 
so the effect of the viscoelasticity 
on the relaxation of the velocity fluctuations (hydrodynamic relaxation) 
was not investigated. 
Here we take the transverse parts of the viscoelastic force and 
the velocity fluctuations into consideration.
We shall then show that the dynamical asymmetry coupling 
between these transverse modes has a strong influence on 
the hydrodynamic relaxation, 
resulting in a wave-number-dependent shear viscosity.
\\

In the present analysis 
we use the two-fluid model\cite{Brochard-deGennes,Brochard-deGennes2,Doi-Onuki} as the 
basic equations describing the dynamics of polymer solutions.
Here we briefly survey the two-fluid model equations. 
The readers who want to know the details of the two-fluid model, 
see the original paper\cite{Doi-Onuki} or review articles
\cite{Tanaka2,Onuki-book}.

Let ${\bv}_p(\br,t)$ and ${\bv}_s(\br,t)$ be the average velocities 
of polymer and solvent, respectively, and $\psi(\br,t)$ be 
the volume fraction of polymer at point $\br$ and time $t$. 
$\psi$ satisfies a conservation law with  ${\bv}_p$:
\EQ
\frac{\partial}{\partial t}\psi =-\nabla\cdot(\psi{\bv}_p).
\label{conservation}
\EN
The volume-averaged velocity $\bv=\psi\bv_p+(1-\psi)\bv_s$ obeys 
the following hydrodynamic equation 
\EQ
\rho\frac{\partial {\bv}}{\partial t}=
-\nabla p+\eta_0\nabla^2{\bv}
-\psi\nabla\frac{\delta F}{\delta \psi}+\nabla\cdot{\hat\sigma}^p,
\EN
where $\rho$ is the average mass density, $\eta_0$ is the solvent viscosity, 
$F$ is the free-energy functional,
${\hat\sigma}^p$ is the viscoelastic stress, and the hydrostatic pressure $p$ is determined by the incompressibility condition
\EQ
\nabla\cdot{\bv}=0.
\EN
For slow motions, the two-fluid model gives the convective velocity 
of polymer $\bv_{p}$ as
\EQ
{\bv}_{p}&=&{\bv}+\frac{(1-\psi)^2}{\zeta(\psi)}
(-\psi\nabla\frac{\delta F}{\delta \psi}+\nabla\cdot{\hat\sigma}^p),\label{polymer-velocity}
\EN
where $\zeta(\psi)$ is the friction coefficient between the two components, 
and is given by
\EQ
\zeta(\psi)=6\pi\eta_0\xi_b^{-2}.
\EN
Here $\xi_b$ is the blob length of order $b/\psi$, with $b$ being the monomer size. 
Equations (\ref{conservation}) and (\ref{polymer-velocity}) imply that the 
nonvanishing viscoelastic stress 
produces a diffusion. This effect is called the stress-diffusion coupling.
The viscoelastic stress due to the network deformations is given by
\EQ
{\hat \sigma}^p=G(\psi)\hW\cdot(\hW-{\hat\delta}).
\EN
This form is derived from the free-energy functional 
introduced in eq. (\ref{F-energy}) shown below. 
Here $G(\psi)$ is the $\psi$-dependent shear modulus, $\hat\delta$ denotes 
the unit matrix and $\hW$ is interpreted  
as being a long-lived strain variable.
We assume that the equation of motion of $\hW$ is 
given by the following dynamic equation\cite{Taniguchi-Onuki,Onuki-book}, 
\EQ
\frac{\partial \hW}{\partial t}+({\bv_p}\cdot\nabla)\hW&=&
(\nabla{\bv}_p^{\dagger}\cdot\hW+\hW\cdot\nabla{\bv}_p)
-\frac{1}{\tau}(\hW-\hat\delta),
\EN
where $\tau(\psi)$ is 
the $\psi$-dependent relaxation time of shear stress.\\

Now we shall investigate the relaxation dynamics of concentration fluctuations
and velocity fluctuations within the linearized approximation of the above set of equations. 
We set $c_0=c(\psi_0)$, $\tau_0=\tau(\psi_0)$ and $G_0=G(\psi_0)$ 
with $\psi_0$ being the average concentration. 
The free-energy functional is assumed 
to be of the following Ginzburg-Landau type 
\EQ
F\{\psi,\hW\}=\int d\br[\frac{1}{2}r_0\psi^2+\frac{1}{2}c_0|\nabla\psi|^2+\frac{1}{4}G_0\sum_{ij}\delta W_{ij}^2],
\label{F-energy}
\EN
where $\delta W_{ij}=W_{ij}-\delta_{ij}$.
The Fourier transform of an arbitary function $g(\br)$ is defined by
\EQ
g_\bk=\int d\br e^{-i\bk\cdot\br}g(\br).
\EN 
In terms of the Fourier components, the linearized equations are given by
\EQ
\rho\frac{\partial\bvt}{\partial t}&=&-\eta_0 k^2\bvt+
i({\hat\delta}-{\hat\bk}{\hat\bk})\cdot\bk\cdot{\hat\sigma}^p_\bk, \label{l1}\\
\frac{\partial}{\partial t}\delta\psi_{\bk}&=&-\Gamma_{\bk}\delta\psi_{\bk}
+\alpha L\bk\cdot\bk\cdot{\hat \sigma}_\bk^p, \label{l2} \\
\frac{\partial}{\partial t}\str_{\bk,ij}&=&-\frac{1}{\tau_0}\str_{\bk,ij}
+iG_0(k_iv_{\bk,j}^{\bot}+k_jv_{\bk,i}^{\bot})\nonumber\\
&&+2G_0\alpha\Gamma_{\bk}{\hat k_i}{\hat k_j}\delta\psi_{\bk}
-G_0\alpha^2Lk^2\sum_{l}({\hat k_i}{\hat k_l}\sigma^p_{\bk,lj}
+{\hat k_j}{\hat k_l}\sigma^p_{\bk,li})\label{l3},\\
\str_{\bk,ij}&=&G_0\delta W_{\bk,ij}\label{l4},
\EN
where $\bvt$ is the transverse part of $\bvk$, $\hat\bk=\bk/k$,
and
\EQ
\alpha=\frac{1}{\psi_0}
\EN
is the dynamical asymmetry parameter of the polymer solutions\cite{Doi-Onuki,Onuki-Taniguchi}.
The kinetic coefficient $L$ is given by
\EQ
L=\psi_0^2(1-\psi_0)^2/\zeta(\psi_0).
\EN
The decay rate in the absence of the viscoelastic coupling is given by
\EQ
\Gamma_{\bk}=L(r_0+c_0k^2)k^2 .
\EN
The viscoelastic force $\bfo$ is defined as
\EQ
\bfo\equiv\nabla\cdot{\hat\sigma}^p.
\EN
For the present purpose, it is convenient to decompose 
the Fourier transform ${\bfo}_\bk$ of the viscoelastic force 
into longitudinal and transverse components,
\EQ
\bfk={\hat{\bk}}{\hat{\bk}}\cdot\bfk+({\hat{\delta}}-{\hat{\bk}}{\hat{\bk}})\cdot\bfk.
\EN
Let us set
\EQ
\bfl\equiv&{\hat{\bk}}{\hat{\bk}}\cdot\bfk,\\
\bft\equiv&({\hat{\delta}}-{\hat{\bk}}{\hat{\bk}})\cdot\bfk.
\EN
From the linearized eqs. (\ref{l1})-(\ref{l4}), we obtain the following two sets of equations of motion for the longitudinal modes and the transverse modes.

For the longitudinal modes, we obtain
\EQ
\frac{\partial}{\partial t}\delta\psi_{\bk}&=&-\Gamma_{\bk}\delta\psi_{\bk}-\alpha LZ_k,{\label{L1}}\\
\frac{\partial}{\partial t}Z_{\bk}&=&-\frac{1}{\tau_0}(1+2\xi_{ve}^2k^2)Z_{\bk}-2G_0\alpha\Gamma_{\bk}k^2\delta\psi_{\bk},{\label{L2}}
\EN
where $Z_{k}=i\bk\cdot\bfl$ and $\xi_{ve}$ is 
a viscoelastic length or a magic length defined by
\EQ
\xi_{ve}\equiv\sqrt{G_0\tau_0\alpha^2L}.
\EN
This length was first introduced 
by Brochard and de Gennes\cite{Brochard-deGennes,Brochard-deGennes2}.
The physical significance of the viscoelastic length has been discussed 
by many researchers (see refs. 10 and 13 for example).

For the transverse modes, we obtain
\EQ
\frac{\partial}{\partial t}\bvt&=&-\frac{\eta_0 k^2}{\rho}{\bvt}+\frac{1}{\rho}\bft,{\label{T1}} \\
\frac{\partial}{\partial t}\bft&=&-\frac{1}{\tau_0}(1+\xi_{ve}^2k^2)\bft-G_0k^2{\bvt}.{\label{T2}}
\EN

From eqs. (\ref{L1}) and (\ref{L2}), the dynamic structure factor  $S_\bk(t)$ is
given by
\EQ
S_{\bk}(t)&=&<\delta\psi_\bk(t)\delta\psi_{-\bk}(0)>\nonumber\\
&=&\frac{S_\bk(0)}{\omega_{+}-\omega_{-}}
\Bigl\{\bigl[\omega_{+}-\frac{1}{\tau_0}(1+2\xi_{ve}^2k^2)\bigr]e^{-\omega_+t}
+\bigl[-\omega_{-}+\frac{1}{\tau_0}(1+2\xi_{ve}^2k^2)\bigr]e^{-\omega_-t}\Bigr
\},\nonumber\\
\EN
where $<\cdots>$ denotes the equilibrium average and
\EQ
\omega_{\pm}=\frac{1}{2}
\Biggl\{\bigl[\Gamma_{\bk}+\frac{1}{\tau_0}(1+2\xi_{ve}^2k^2)\bigr]
\pm{\sqrt{\bigl[\Gamma_{\bk}+\frac{1}{\tau_0}(1+2\xi_{ve}^2k^2)\bigr]^2-4\frac{\Gamma_\bk}{\tau_0}}}\Biggr\}.
\EN
This form of the dynamic structure factor was first proposed by 
Brochard and de Gennes\cite{Brochard-deGennes,Brochard-deGennes2} 
on the basis of a phenomenological argument.
Afterwards, Doi and Onuki\cite{Doi-Onuki} 
derived the same form of the structure factor by analysing 
their two-fluid model equations which is employed in the present analysis 
(see also ref. 11).\\

The viscoelasticity is characterized by the complex shear modulus 
$G^{*}_\bk(\omega)=G'_\bk(\omega)+iG''_{\bk}(\omega)$, 
where $G'_\bk(\omega)$ and $G''_\bk(\omega)$ are 
called the storage modulus and the loss modulus, 
respectively\cite{Doi-Edwards}.
Since eqs. (\ref{T1}) and (\ref{T2}) are linear, we may assume, without loss of generality, that $\bvt(t)$ and $\bft(t)$ depend on time as 
\EQ
\bvt(t)&=&\bvt e^{i\omega t},\\
\bft(t)&=&\bft e^{i\omega t}.
\EN  
From eqs. (\ref{T1}) and (\ref{T2}), 
the total shear stress ${\hat\sigma}_\bk(t)$ is expressed as
\EQ
{\hat\sigma}_{\bk}(t)=\frac{G^{*}_\bk(\omega)}{i\omega}\bigl\{i\bk\bvt+i[\bk \bvt]^\dagger\bigr\}e^{i\omega t},
\EN
where $[\bk \bvt_\omega]^\dagger$ is the transpose of 
the dyadic $\bk \bvt_\omega$. 
In the present system, $G'_\bk(\omega)$ and $G''_\bk(\omega)$ are given by
\EQ
G'_{\bk}(\omega)&=&\frac{G_0\tau_0^2\omega^2}{\omega^2\tau_0^2+(1+\xi_{ve}^2k^2)^2},\label{storage}\\
G''_{\bk}(\omega)&=&\omega\bigl[\eta_0+G_0\tau_0\frac{1+\xi_{ve}^2k^2}{\omega^2\tau_0^2+(1+\xi_{ve}^2k^2)^2}\bigr],\label{loss}
\EN
which are non-local in both space and time. 
The complex shear viscosity is given by
\EQ
\eta_\bk(\omega)=\frac{G^{*}_\bk(\omega)}{i\omega}.
\EN
The imaginary part of $G^{*}_\bk(\omega)$ gives the shear viscosity 
for long-time scale motion $(\omega\cong 0)$ as 
\EQ
\eta_\bk(0)=\lim_{\omega \to 0}\frac{G''_\bk(\omega)}{\omega}=\eta_0+\eta_p\frac{1}{1+\xi_{ve}^2k^2}, \label{shear-viscosity-das}
\EN
where we have defined $\eta_p\equiv G_0\tau_0$. 
The asymptotic behavior of $\eta_\bk(0)$ is represented by
\EQ
\eta_\bk(0)&\cong&\eta_0+\eta_p~~~~(k\xi_{ve}\ll 1),\\
&\cong&\eta_0+\eta_p\frac{1}{\xi_{ve}^2k^2}~~~~(k\xi_{ve}\gg 1)\label{shear-2}.
\EN 
For polymer solutions, the two-fluid model gives the viscoelastic length as\cite{Doi-Onuki,Onuki-book} 
\EQ
\xi_{ve}\cong (\eta_p/\eta_0)^{\frac{1}{2}}\xi_{b}.\label{ve-blob}
\EN
Thus, eq. (\ref{shear-2}) is also expressed as
\EQ
\eta_\bk(0)&\cong&\eta_0\bigl(1+\frac{1}{\xi_b^2k^2}\bigr),~~~~(k\xi_{ve} \gg 1),
\EN
which explicitly shows that the hydrodynamic interaction becomes weak 
beyond the blob length $\xi_{b}$.
It is worth mentioning that this property is due to 
the dynamical asymmetry coupling ($\alpha\ne 0$) 
between the viscoelastic stress and the velocity fluctuations.
For polymer solutions, it is well known that the diffusion process is 
drastically influenced by the dynamical asymmetry coupling
between the longitudinal modes, namely the so-called stress-diffusion coupling\cite{Brochard-deGennes,Brochard-deGennes2,Doi-Onuki}. 
The present analysis shows that 
the dynamical asymmetry coupling between 
the transverse modes is also important for 
the hydrodynamic relaxation process.  
Previously Edwards and Freed also discussed 
the same effect for polymer solutions by a microscopic arguments 
\cite{Edwards-Freed,Edwards-Freed2,Doi-Edwards}.
However, they considered a system where polymers are assumed not to be entangled. The present result is due to the collective motion on a scale of the viscoelastic length, $\xi_{ve}$, resulting from the entanglement of the polymers.\\

We now investigate a more general situation. 
Within the linear regime, 
the total shear stress ${\hat\sigma}_\bk(t)$ can be written as
\EQ
{\hat\sigma}_\bk(t)=\eta_0\hat\kappa_\bk(t)+\int_{-\infty}^tdt'G^{(p)}_\bk(t-t')\hat\kappa_\bk(t'),\label{shear-relaxation}
\EN
where 
\EQ
{\hat\kappa}_\bk(t)=i\bk\bvt(t)+i[\bk\bvt(t)]^\dagger.
\EN
$G^{(p)}_\bk(t)$ is called the shear relaxation modulus 
and fulfills the following relation\cite{Doi-Edwards} 
\EQ
\frac{G^{*}_\bk(\omega)}{i\omega}=\eta_0+\int_0^\infty dt G_\bk^{(p)}(t)e^{-i\omega t}.\label{relation}
\EN
With the aid of eqs. (\ref{storage}) and (\ref{loss}), 
the inverse Laplace transformation of eq. (\ref{relation}) gives
\EQ
G^{(p)}_\bk(t)=\frac{\eta_p}{\tau_0}\exp\biggl[-\frac{1+\xi_{ve}^2k^2}{\tau_0}t\biggr].
\EN
It should be noted that 
the relaxation time is given not by $\tau_0$ but by $\tau_0/[1+(\xi_{ve}k)^2]$.

We can rewrite eq. (\ref{shear-relaxation}) in real space as
\EQ
{\hat\sigma}(\br,t)=\eta_0\hat\kappa(\br,t)+\int_{-\infty}^tdt'\int d\br' G^{(p)}(t-t',\br-\br')\hat\kappa(\br',t'),\label{stress}
\EN
where
\EQ
\hat\kappa(\br,t)=\nabla\bv^\bot(\br,t)+[\nabla\bv^\bot(\br,t)]^\dagger,
\EN
and $G^{(p)}(t-t',\br-\br')$ is given by
\EQ
G^{(p)}(t-t',\br-\br')=\frac{\eta_p}{\tau_0}\Biggl[\frac{\tau_0}{4\pi\xi_{ve}^2(t-t')}\Biggr]^{\frac{3}{2}}\exp\biggl[-\frac{(t-t')}{\tau_0}-\frac{|\br-\br'|^2}{4\xi^2_{ve}}\frac{\tau_0}{(t-t')}\biggr].\label{propagator}
\EN
Equations (\ref{stress}) and (\ref{propagator}) explicitly show that 
the shear stress $\hat\sigma(\br,t)$ is dominated by the velocity fluctuations 
of the distance $|\br-\br'|\sim O(\xi_{ve})$ and of the time $t-t'\sim O(\tau_0)$.
If we take the limit $\xi_{ve}\rightarrow 0$, eq. (\ref{stress}) reduces to
the well-known form as follows:
\EQ
{\hat\sigma}(\br,t)=\eta_0\hat\kappa(\br,t)+\frac{\eta_p}{\tau_0}\int_{-\infty}^tdt' \exp\biggl[-\frac{(t-t')}{\tau_0}\biggr]\hat\kappa(\br,t').
\EN\\

In this study, we have investigated the relaxation dynamics of the
concentration fluctuations and the velocity fluctuations 
by the linearized two-fluid model equations.
The present results can be applied to the polymer solutions in a one-phase 
region far from the critical point. 
However, in the vicinity of the critical point 
the nonlinear hydrodynamic interaction, 
resulting from the streaming type mode coupling, becomes dominant
\cite{Tanaka-Nakanishi-Takubo}.
Taking such nonlinear hydrodynamic interaction into account
by means of the mode coupling theory,
we will report 
the critical dynamics of polymer solutions elsewhere\cite{Furukawa}.\\

The present author wishes to thank Professor Akira Onuki for useful comment. 
This work was completed during his visit to Hiroshima University where 
the author had much to be thankful to 
Professor Takao Ohta and Dr. Tohru Okuzono for their helpful discussions.

\end{document}